# Mapping optical, chemical, structural features in $ZrO_2$ via cross-sectional SEM-Cathodoluminescence correlation microscopy

Ricardo Vidrio[1*], Yuhan Tong[1*], Junliang Liu[4], Bil Schneider[5], William O. Nachlas[5], Nathan Curtis[2], Maryam Zahedian[1], Alexander Kvit[3], Hongliang Zhang[3], Zefeng Yu[2], Ximeng Wang[2], Yongfeng Zhang[2], Adrien Couet[2], and Jennifer T. Choy[1]

[1]Department of Electrical and Computer Engineering, University of Wisconsin-Madison, United States

[2]Department of Nuclear Engineering and Engineering Physics, University of Wisconsin-Madison, United States

[3]Department of Material Science and Engineering, University of Wisconsin-Madison, United States

[4]Department of Materials Science and Engineering, FAMU-FSU College of Engineering, Florida State University, United States

[5]Department of Geoscience, University of Wisconsin-Madison, United States

* These authors contributed equally to this work.

## Abstract

Understanding how nanoscale heterogeneities influence charge transport and mass transfer in oxides is critical for developing advanced materials for energy and electronic uses. In high-temperature applications, the formation of thermal oxides with complex chemical and structural features plays a central role in material lifetime. While thermally grown zirconia ($ZrO_2$) on zirconium alloys exhibits strong chemical and microstructural gradients across the oxide thickness, linking these heterogeneities to electronic-defect landscapes remains challenging. We demonstrate cross-sectional scanning electron microscope-cathodoluminescence (SEM-CL) as a mesoscale probe of spatial variations in luminescence in zirconia and establish correlations with co-registered electron backscatter diffraction (EBSD) and electron probe micro-analysis (EPMA) on the same region. The SEM-CL signal is dominated by the ~2.7 eV defect band, but its intensity varies strongly across the oxide cross section. Correlative EBSD-CL analysis reveals that CL intensity increases with grain area and decreases at the grain boundaries, consistent with enhanced non-radiative recombination associated with microstructural disorder. EPMA mapping shows that a substantial fraction of CL-dark features co-localize with secondary phase precipitates enriched in iron. These results show that SEM-CL contrast in corrosion-grown $ZrO_2$ is controlled by both chemical heterogeneity and microstructural disorder, underscoring the need for correlative registration to interpret CL images. This multi-modal approach provides an efficient route to connect electronic properties and luminescence signatures across complex oxide cross sections to underlying chemistry and microstructure, thereby providing a pathway to relate local defect landscapes to regions likely to bias electronic/ionic transport during oxidation.

## 1. Introduction

High-temperature oxides are critical in a wide range of energy technologies, from protective coatings in gas turbines to corrosion barriers in nuclear reactors, and corrosion itself carries an enormous economic burden, estimated at 3.4% of the global GDP or roughly $2.5 trillion USD [1]. In these applications, oxides often form or operate under extreme environments involving high temperatures, mechanical stress, radiation exposure, and chemical gradients [2]. Rather than being one homogeneous phase, these oxides frequently exhibit complex microstructures and nanoscale chemical heterogeneities that strongly influence their electronic and ionic transport. Zirconium dioxide ($ZrO_2$) is a prominent example: it forms during corrosion of zirconium-alloy nuclear fuel cladding in water-moderated nuclear reactors, where the protective oxide scale it provides governs corrosion kinetics. Studying the local electronic and chemical properties of $ZrO_2$ formed under reactor-relevant conditions provides valuable insights not only into corrosion mechanisms but also into the broader principles governing transport phenomena in heterogeneous oxide systems.

Directly probing micro- and nanoscale variations in electronic structure and charge transport is generally challenging. Luminescence offers a practical window: the intensity and spectral content of $ZrO_2$ emission, whether excited optically, thermally, or by electrons, reveal the electronic structure and defect densities in the oxide [3] [4] [5] [6] [7] [8] [9]. Spectroscopic studies have established that $ZrO_2$ luminescence primarily originates primarily from intrinsic defects, particularly oxygen vacancies, with characteristic emission at ~2.4-2.7 eV [9] [10] [11] [12]. In this work, we use scanning electron microscope cathodoluminescence (SEM-CL) as a mesoscale probe of this defect landscape: we map cross-sectional variations in luminescence across thermally grown $ZrO_2$ on a zirconium alloy and correlate them with underlying microstructure and chemistry through co-registered electron backscatter diffraction (EBSD) [13] and electron probe micro-Analysis (EPMA) [14]. Because the oxide is probed in the cross-section, these correlations are resolved along the direction of oxide growth.

SEM-CL has a long track record in other materials communities. In semiconductors, it is used to map electronic structure, defect densities, and impurity distributions at high spatial resolution [15] [16] [17] [18] [19] [20], because CL intensity and spectra are governed by carrier recombination dynamics and therefore resolve the local defect landscape and recombination centers. In the geosciences, correlative measurements with elemental analysis have demonstrated sensitivity to both trace chemistry and microstructure. In quartz, trace titanium and aluminum act as CL activators [21] [22] [23] while iron (Fe) acts as a quencher [24] [25] [26] [27], and correlating SEM-CL with Wavelength-Dispersive X-ray Spectroscopy (WDS) maps reveals trace-element distributions at the nanometer scale [21]. SEM-CL imaging in quartz has likewise been used to visualize sub-grain boundaries and twinning structures, whose associated dislocation-related defects act as active CL centers [28]. This combination of electronic, chemical, and microstructural sensitivity is well suited to corrosion-grown $ZrO_2$, where grain boundaries, impurity segregation, and second-phase precipitates (SPPs) are common [29] [30] [31].

In corroded zirconium alloys, Yueh and Cox pioneered SEM-CL imaging of the spatial distribution of SPPs and showed that CL contrast correlates with the presence of SPPs [3] [4]. However, these SEM-CL results were interpreted qualitatively by visual correspondence to microstructural/precipitate features. A key ambiguity therefore remains: in corrosion-grown $ZrO_2$, CL contrast convolves radiative emission, chemical quenching/activation, and non-radiative recombination, and cannot be interpreted without independent structural and chemical context. The central question here is thus: *to what extent can spatial variations in the cross-sectional CL signal be attributed to local chemistry vs local microstructure in thermally grown zirconia on zirconium alloys?*

To answer this question, we co-registered SEM-CL, EPMA/WDS elemental mapping, and EBSD on the same oxide cross section. A multimodal registration workflow aligns EBSD and CL maps from an identical region, enabling correlation between CL intensity and local microstructure, and a machine-learning-assisted segmentation step separates discrete CL-dark features (candidate precipitate-associated regions) from broader background variations for objective, semi-quantitative comparison with EPMA chemistry. Using these co-registered datasets, we show that CL intensity is systematically suppressed in (i) Fe-enriched regions identified by EPMA and (ii) microstructurally disordered, grain-boundary-rich regions identified by EBSD, demonstrating that SEM-CL provides a sensitive mesoscale map of electronic defects with chemical and structural origins. The primary contribution of this work is methodological: a workflow that separates chemical from microstructural contributions to CL contrast within a single oxide cross-section. Because distance from the metal/oxide interface encodes oxide age, precipitate oxidation state, and grain morphology, the thermally grown oxide is itself an internally graded system, and the correlations reported here are examined across this natural gradient. Extending the workflow to a parametric study across alloy compositions and oxidation conditions is a natural direction for future work.

## 2. Methods

The material investigated in this work was a recrystallized Zr-Fe-Cr model alloy (Zr-0.4 wt% Fe-0.2 wt% Cr) fabricated by Westinghouse LLC, by vacuum arc melting (re-melted four times for homogeneity), followed by hot rolling, pickling, and multiple cycles of cold rolling and vacuum annealing to produce about 0.8-mm-thick strip. The sample was processed at 720 °C selected to promote precipitate coarsening. Synchrotron XRD indicates that the SPP index as $Zr(Cr,Fe)_2$ C14 (hcp Laves) [32]. In the ZrFeCr model alloy family, precipitate size is primarily controlled by processing temperature while precipitate volume fraction increases with total Fe+Cr content. For this alloy, the SPP volume fraction was determined to be about 1% and the average SPP diameter varies between 35 and 110 nm. The sample was oxidized in (static) air at 700 °C for 36 hours in a box furnace to grow a relatively thick $ZrO_2$ scale. The oxide thickness was inferred from weight gain and confirmed by cross-sectional SEM, yielding a thickness of approximately 30 µm. Cross-sections were prepared by low-speed saw sectioning, mounting in epoxy, and mechanical polishing to a final 0.5 µm colloidal silica finish to enable high-quality imaging. The zirconium alloy was then placed within epoxy to allow for easy handling during imaging. Three different electron microscopes were used throughout the experiment, each with its own capability for providing an electron, X-ray, or optical photon imaging technique. The key experimental method in this work is SEM-based cathodoluminescence (SEM-CL), in which the SEM electron beam generates electron-hole pairs in the oxide and the subsequent radiative recombination produces photons whose intensity and spectrum report on defect- and impurity-related electronic states. As these carriers recombine, they emit photons whose energies are sensitive to both extrinsic and intrinsic defects within the material. Extrinsic defects arise from the incorporation of foreign cations and anions into the lattice [21], whereas intrinsic defects include structural imperfections or deviations in stoichiometry [33].

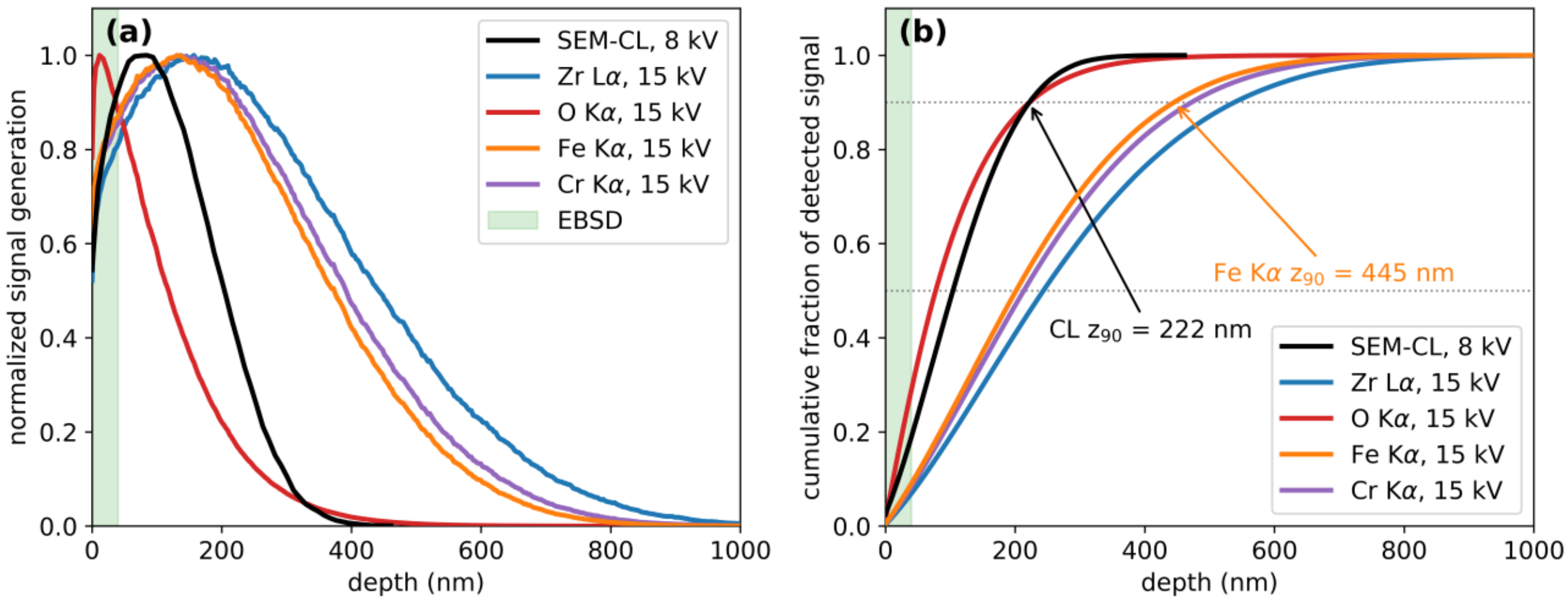

Figure 1: Monte Carlo simulation of signal sampling depths in $ZrO_2$ at the operating conditions of this study (see Methods). (a) Normalized depth profiles of SEM-CL generation (8 kV) and emitted EPMA X-ray intensity (15 kV; Zr Lα, O Kα, Fe Kα, Cr Kα, absorption-corrected for the 40 deg take-off angle). Shaded band: EBSD information depth (~10-40 nm, Kikuchi-limited *[34]*). (b) Cumulative depth distributions; dotted lines mark the 50% and 90% levels, and $z_{90}$ values are annotated for SEM-CL (222 nm) and Fe Kα (445 nm).

A Hitachi S-3400N Type-II variable pressure SEM was used for secondary electron imaging, an attached Gatan PanaCL CL detection system was used for gathering CL images from the sample. The Gatan CL detection system comes equipped with three different bandpass color filters that allow for viewing of the optical wavelengths of red, green, and blue light that emanate from the sample. The PanaCL detector has a range of 185-850 nm. The red filter allows wavelengths of 595-850 nm; The green filter allows 495-575 nm and the blue allows 185-510 nm. Spectrally resolved CL was not acquired, as this requires a monochromated detection system. Spectral assignments in this work are therefore based on published spectroscopic measurements of monoclinic $ZrO_2$ [11]. Throughout analysis of the sample, 8 kV of accelerating voltage was used for both SE and CL image gathering.

EBSD data were acquired at an accelerating voltage of 10 kV using a Gemini 300 SEM at the Nanoscale Imaging and Analysis Center (NIAC at the Wisconsin Centers for Nanoscale Technology) at the University of Wisconsin-Madison, equipped with a CMOS-based EBSD detector of Oxford Instruments Symmetry S3 for high-sensitivity collection of Kikuchi patterns generated by backscattered electrons.

Finally, a Cameca SXFive Field Emission Electron Probe Microanalyzer (FE-EPMA) equipped with five wavelength dispersive spectrometers (WDS) was utilized for elemental mapping of the sample, consisting of Zr, Fe, Si, Cr, and O. WDS maps were acquired by rastering the electron beam across a fixed area of the sample while acquiring X-ray counts on five spectrometers consecutively. The primary beam was accelerated at the sample with 15-kV voltage and 0.3-μA beam current with a fully focused (~60 nm) beam size. WDS maps were quantified using the Mean Atomic Number background technique [35] using a calibration curve generated from analysis of 25 high purity metal standards.

Since SEM-CL, EBSD, and EPMA acquire signal from different characteristic depths, we performed Monte Carlo simulations of electron trajectories in monoclinic $ZrO_2$ (CASINO v2.5 [36]) to quantify the sampling depth of each modality at the operating conditions of this study. Fe and Cr were included as dilute tracers at their alloy-composition weight fractions. This approach gives the depth distribution of the signal averaged over the interaction volume; local composition within individual precipitates is not resolved by this treatment. Simulations used 50,000 primary electrons incident on a semi-infinite $ZrO_2$ target ($\rho$ = 5.68 g $cm^{-3}$), with beam diameters of 10 nm at 8 kV (SEM-CL conditions) and 60 nm at 15 kV (EPMA conditions), matching the respective instruments. The SEM-CL sampling profile was taken as proportional to the depth distribution of deposited energy, i.e., the electron–hole pair generation rate; self-absorption of the emitted light was neglected, as the CL bands of interest (~2.0-2.7 eV) lie well below the $ZrO_2$ bandgap (~5.8 eV), for which the bulk oxide matrix is essentially transparent. For EPMA, the X-ray generation profiles of the mapped lines (Zr Lα, O Kα, Fe Kα, Cr Kα) were computed and corrected for X-ray absorption along the emission path at the 40-deg spectrometer take-off angle, yielding the emitted-intensity depth profiles that determine the detected signal. The EBSD information depth is not governed by the interaction volume but by the loss of Kikuchi-pattern coherence upon inelastic scattering and is taken from the literature as ~10-40 nm [34]. The simulation results establish the depth hierarchy and frame the interpretation of co-localization and mismatch between the correlated maps in Section 3.

The two main computational methods used to discern the data and microscopy images were (i) Reactiv'IP® IPSDK software, and (ii) image registration techniques. IPSDK was used to train a machine-learning segmentation model to detect the dark, approximately circular features in SEM-CL images and enable correlation of these features with corresponding regions in elemental WDS maps.

3. Results

### *3.1 SE and SEM-CL image comparisons*

Shown on Figure 2 are SE and SEM-CL images of identical regions on the $ZrO_2$ cross section. Platinum crosses were deposited on the sample by Focused Ion Beam (FIB) (NIAC) to enable accurate cross-correlation/registration between images, and as a marker to facilitate rapid identification and relocation of the region of interest. Whereas only surface features and coarse structure are visible on the SE image (Figure 2a), Figure 2b shows spatially varying CL features emanating from the sample when 8 kV electrons are impingent on the $ZrO_2$ sample. The CL is constrained to only the oxide layer and decreases sharply at both the metal-oxide and oxide-epoxy boundary regions. At the magnification and contrast settings shown, the oxide exhibits no obvious through-thickness CL gradient, although local intensity variations are apparent. This qualitative behavior is consistent with prior findings by Yueh et. al [4] [3].

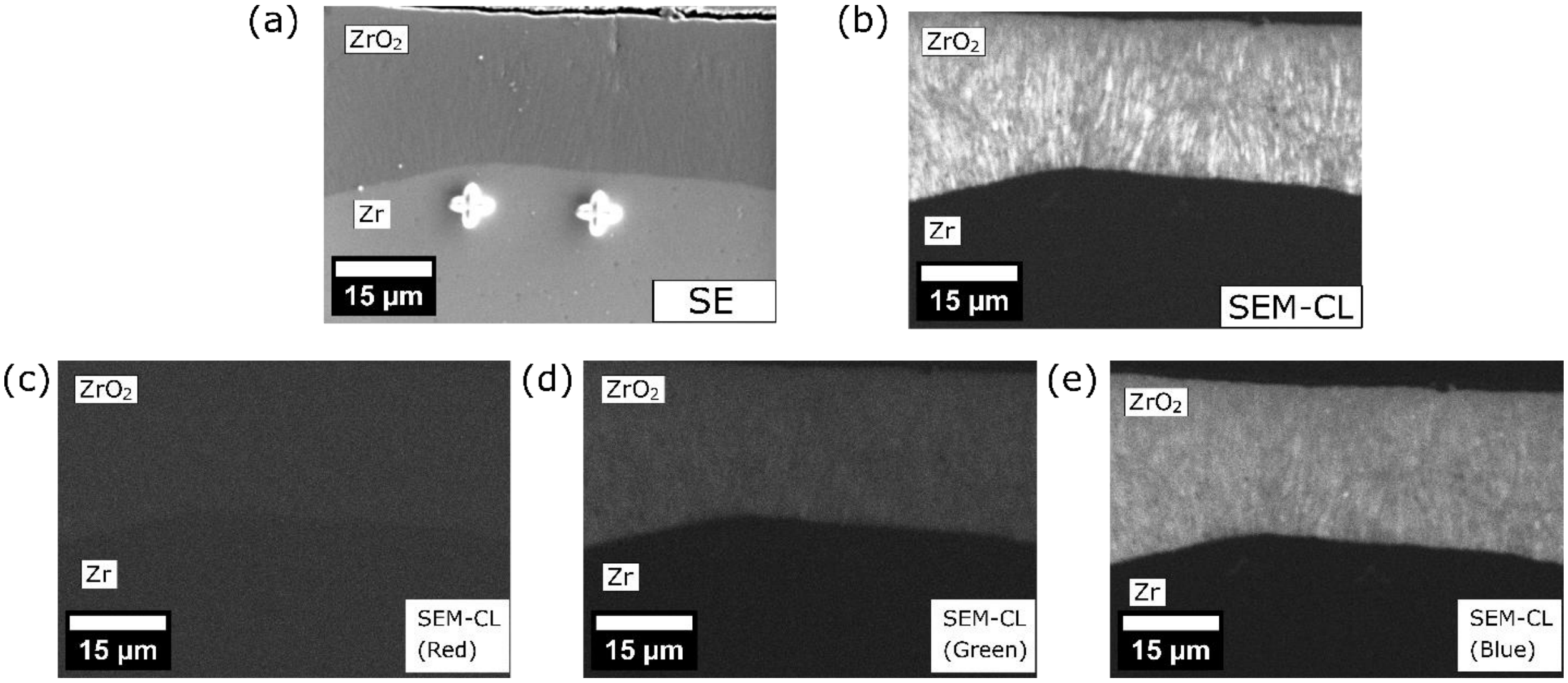

Figure 2: (a) Secondary Electron (SE) image showing surface topography. (b) Unfiltered SEM-CL image of $ZrO_2$ sample (c) Red (d) Green (e) Blue filtered SEM-CL images of $ZrO_2$ sample from the same region.

Spectrally filtered SEM-CL images collected from the oxidized zirconium sample are shown on Figure 2c-e. The luminescence is most prominent in the blue band, with moderate emission in green and almost no luminescence in the red. This observation is consistent with a broad CL spectrum that peaks in the blue wavelengths, and is consistent with prior computational and experimental work that has identified the source of $ZrO_2$ luminescence to oxygen vacancies ($F^+$ center) with a characteristic emission centered around 2.7 eV (corresponding to ~460 nm) [7] [9] [10]. The moderate emission in the green channel can be assigned to oxygen-vacancy-related centers. In yttria-stabilized $ZrO_2$, sub-bands at ~550 nm and ~600 nm have been attributed to $F_A^+$ and $F_{AA}^+$ centers, respectively referring to oxygen vacancies with one and two substitutional cations as nearest neighbors. Emission from these centers is progressively red-shifted from the intrinsic $F^+$ band at 460 nm [7]. First-principles calculations on nominally pure monoclinic $ZrO_2$ similarly assign the dominant ~500 nm emission not to the isolated vacancy, whose calculated transition at 2.81 eV is consistent with the observed band, but to a Ti impurity adjacent to oxygen vacancy complex ($e_g \rightarrow t_{2g}$) or $F_A$ center ($F_A^0 \rightarrow F_A^+$) [37]. In our thermally grown oxide, an analogous perturbation of oxygen vacancies by neighboring Fe or Cr dissolved dilutely in the $ZrO_2$ lattice [38] provides a natural origin for the green-channel signal. The near absence of red emission is likewise consistent: at the dilute Fe or Cr concentrations in the oxide, doubly perturbed $F_{AA}^+$ centers should be rare.

Overall, the SEM-CL image reveals largely uniform broadband visible luminescence across the oxide, strongest in the blue channel, superimposed with spatial variations that include banded striations and discrete dark spots. These features indicate local changes in the defect landscape and are further investigated with EBSD and EPMA for their connection with underlying microstructural heterogeneity and localized chemical variations.

### *3.2 EBSD and SEM-CL comparisons*

To link crystallographic texture to luminescence in the oxide, we performed correlative EBSD and SEM-CL on the same cross-sectional region of $ZrO_2$. EBSD maps were acquired on a polished surface and registered to the SEM-CL image using surface features and the metal/oxide boundary as fiducials. We utilized the multimodal reconstruction framework developed by Charpagne et al. [39] to correct for the relative transformation such as translation, rescaling, and rotation between the EBSD and SEM-CL map overlay. We initially evaluated their numerical optimization method based on the Covariance Matrix Adaptation–Evolution Strategy (CMA-ES), which has previously been demonstrated for SE↔EBSD alignment [39]. However, although CMA-ES-based nonrigid warping has been demonstrated for EBSD↔SE alignment, we found that applying the same approach to EBSD↔SEM-CL alignment introduced non-physical distortions without improving registration quality. We attribute this to modality differences: while SE and EBSD share similar grain-boundary contrast, SEM-CL contrast is governed by local defect recombination and does not consistently reproduce the EBSD-visible microstructure. In addition, charging-induced drift in the insulating $ZrO_2$ cross-section introduces line-based shear distortions that are not well modeled by smooth polynomial warping. Therefore, for EBSD↔SEM-CL registration we used only the affine transformation determined from feature correspondences. In this case, it preserves geometric fidelity without introducing artifacts.

Figures 3a-b show co-localized SEM-CL and EBSD images without reconstruction and highlight that the CL contrast is strongly structured on the scale of the microstructure. The SEM-CL image exhibits a relatively uniform blue luminescence background with pronounced banded/striated intensity variations and occasional dark spots, while the EBSD map shows corresponding spatial changes in grain morphology and boundary density. After multimodal image registration the EBSD overlays reveal that CL striations align with ones of differing grain structure: regions containing elongated and thin grains with higher grain-boundary density appear systematically dimmer in CL, whereas coarser-grained regions appear brighter (select regions of interest (ROIs) are shown in Figure 3c).

To quantify this relationship, we computed the average grain area and mean orientation spread from EBSD across multiple ROIs (see Supplementary Information for grid definition and analysis) and plotted it against the mean SEM-CL intensity from the corresponding regions of a single SEM-CL image. Mean orientation spread refers to the average misorientation angle between each pixel in a grain and that specific grain's overall average orientation. Figure 3d and 3e reveal an overall positive correlation between CL intensity and average grain area and negative trend for CL intensity with mean orientation spread. One data point in Figure 3f-g deviates from the general trend, exhibiting an exceptionally high mean CL relative to its average grain area. We hypothesize that this discrepancy arises from the larger information depth of SEM-CL signal compared to the more surface-limited EBSD measurement. Specifically, the CL signal may capture high-intensity monoclinic grains beneath the cross-sectional surface that are not detected in the EBSD map. Overall, the observed trends are consistent with enhanced non-radiative recombination associated with grain boundaries and microstructural disorder, which lowers the local radiative recombination probability and suppresses the local CL yield relative to more crystalline, larger-grain regions.

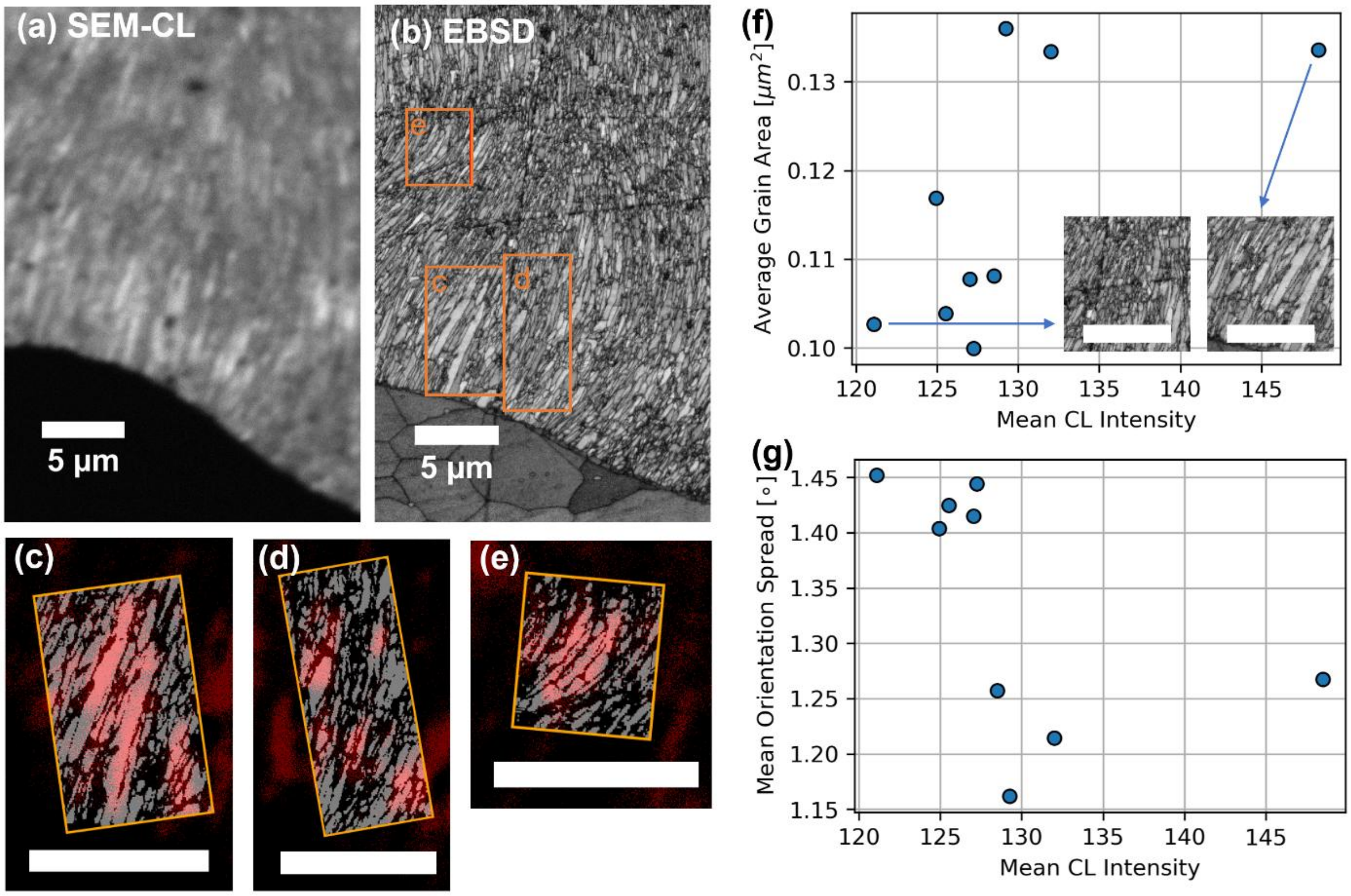

Figure 3: (a) SEM-CL and (b) EBSD maps from a polished cross section of $ZrO_2$ showing intensity modulations (striations) in the SEM-CL image and lamellar/columnar microstructure in the EBSD image near the metal/oxide interface. Colored boxes outline some of the regions of interest (ROIs) used for correlative analysis. (c)-(e) Examples of EBSD-CL co-registration for selected ROIs: EBSD (white) grain information (inset) mapped onto the corresponding SEM-CL (red) intensity field; the orange outline denotes the transformed EBSD footprint after multimodal registration. Scalebars correspond to 5 $\mu$m. (f)-(g) Quantitative comparison across 9 (3x3 grid of the same size) ROIs showing that mean CL intensity trends positively with EBSD-derived average grain area and negatively with average grain mean orientation spread (a measure of intragranular misorientation). These results indicate that brighter CL regions tend to correspond to coarser-grained microstructure with lower density of grain boundaries, and that disorder within a grain suppresses CL. Scalebars correspond to 5 $\mu$m.

### *3.3 EPMA and SEM-CL comparisons*

SEM-CL images of the oxide cross sections (Figures 2b and 3a) reveal dark, ~1-μm features within 15 μm of the metal-oxide layer boundary. Those dark features represent specific CL quenching features with the thermally grown $ZrO_2$. To probe the origin of CL quenching, we performed elemental mapping using EPMA. The Zr (wt%), O (at%), Fe (at%), and Cr (at%) maps are shown alongside their respective scale bars to improve image contrast and better visualize SPP distribution in oxide (Figure 4). Platinum crosses deposited beneath the oxide are visible on the Zr and O maps and used for co-registration.

The oxide is well defined by the O map, allowing the clear identification of the metal/oxide interface. From the Zr and O elemental maps, one can discern regions in the oxide layer with lower amounts of both Zr and O corresponding to higher concentrations of both Fe and Cr. Notably, the dark specks in the O map (indicating O-depleted regions) correspond to Fe-Cr rich particles in the inner oxide, whereas these features are absent in the outer oxide layer. This observation is consistent with the well-documented oxidation behavior of $Zr(Fe,Cr)_2$ C14 Laves phase SPPs [40] [41]. As the oxide layer grows into the alloy substrate following the $ZrO_2$ n-type oxidation behavior, the $Zr(Fe,Cr)_2$ SPPs become embedded in the oxide and undergo delayed oxidation relative to the Zr matrix. Near the metal/oxide interface, the Fe-Cr rich particles initially remain metallic, which accounts for the O-depleted contrast observed in the O maps. As the oxide continues to grow and SPPs become further from the interface, they begin to oxidize. During this process, Fe diffuses out of the precipitates while Cr remains and oxidizes locally, leaving Cr-O rich remnants [42] [41] [43]. The resulting Cr-rich precipitates appear relatively homogenously distributed across the oxide, whereas Fe-rich precipitates are confined to the inner region near the metal/oxide interface. This progressive oxidation of Fe has been directly observed by microbeam synchrotron X-ray absorption near-edge spectroscopy in oxide layers of the same alloy class, which shows metallic Fe near the metal/oxide interface developing more oxidation with increasing distance into the oxide [44].

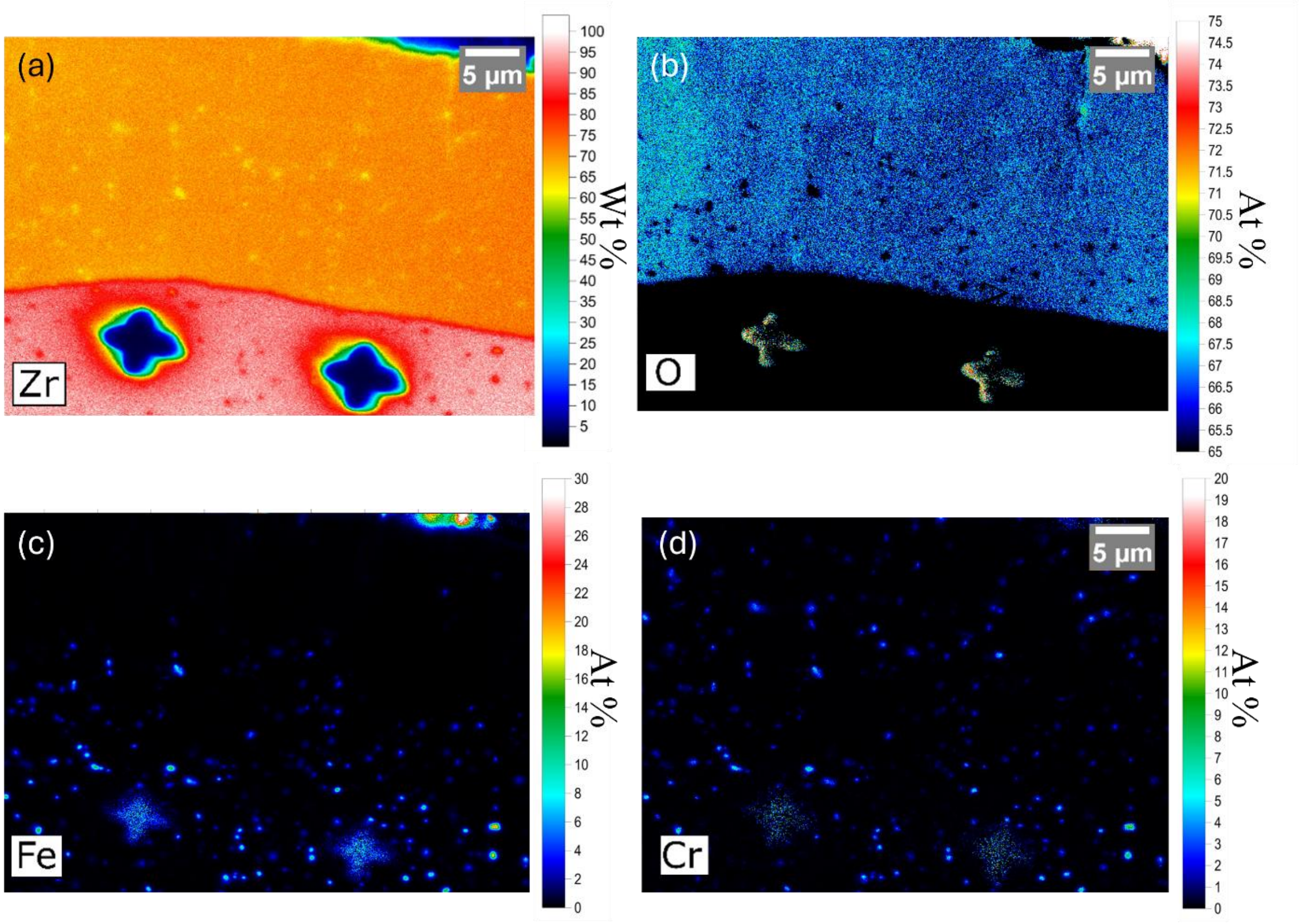

Figure 4: WDS elemental maps performed on corroded zirconium alloy sample. (a) Zr in wt%, and O map is thresholded to the physically relevant 65–75 at% range, which improves intra-oxide contrast and reveals Fe-Cr-rich particles as O-depleted features in the oxide layer (b) O in at% (c) Fe in at% (d) Cr in at%. Platinum crosses used for co-registration are visible in bottom.

Figure 5 compares the EPMA map (Fe overlaid on SE and Zr images) with a co-located SEM-CL image. Co-registration was done using the platinum crosses and morphological landmarks. Numbered markers indicate pink sites with direct co-localization of Fe precipitates and CL-dark features, and cyan sites where there are small spatial offsets ≤ 2 µm, consistent with registration uncertainty and differing interaction volumes between SEM-CL and EPMA (primarily due to CL being acquired at lower accelerating voltages). A perfect 1:1 correspondence is not expected because the two methods probe different sampling depths and contrast mechanisms: EPMA provides an elemental map integrated over a deeper sampling volume ($z_{90} \approx 445$ nm for Fe Kα; Figure 1), while SEM-CL reports the local radiative response of a shallower excited volume ($z_{90} \approx 220$ nm) and is additionally sensitive to defect- and microstructure-driven nonradiative recombination. These two asymmetries account for the two types of mismatches: Fe precipitates without corresponding CL-dark spots are consistent with Fe residing within the deeper EPMA sampling volume but below the CL generation depth, while CL-dark features without an Fe counterpart point to nonradiative recombination of microstructural or non-Fe defect origin [2].

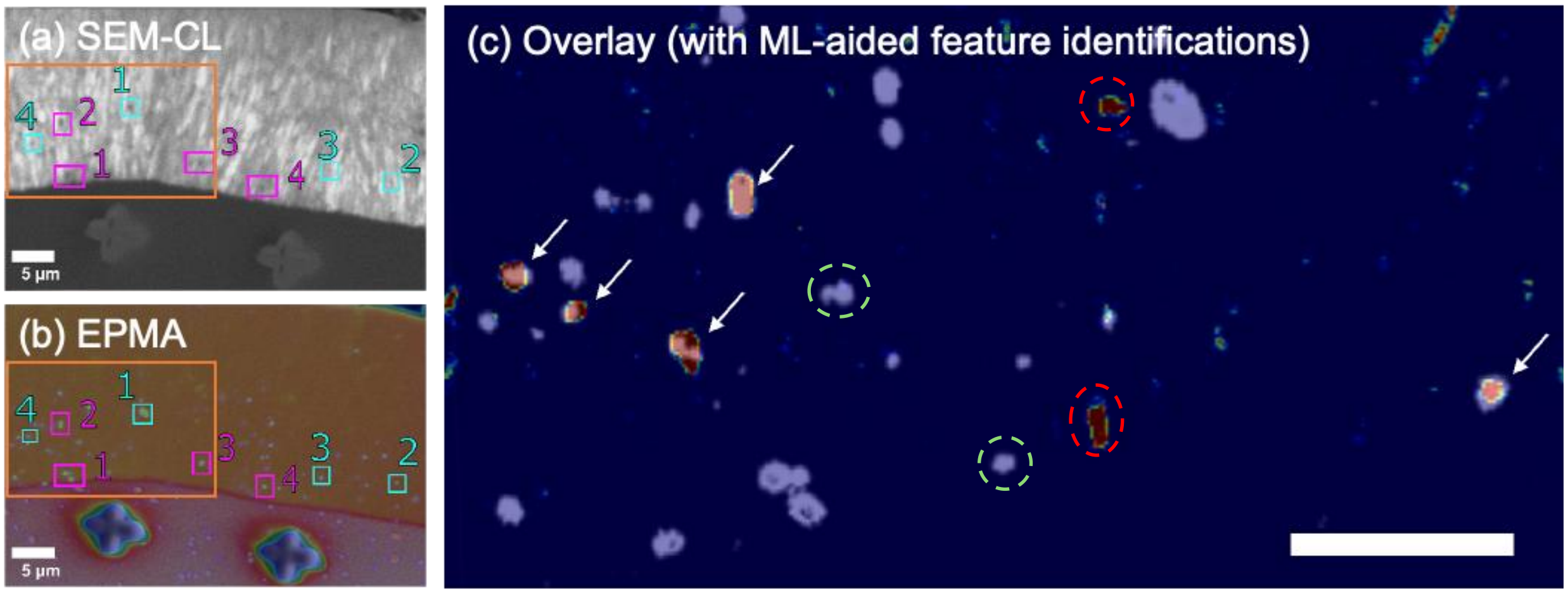

Figure 5: Cross-correlation of Fe EPMA with SEM-CL. (a) SEM-CL image zoomed into the approximate region that was imaged on the EPMA, shown in (b). The labeled regions are ones where Fe precipitates correspond to either a well-aligned or shifted (within 2 $\mu m$) dark region in SEM-CL. The orange boxes in both figures indicate the selected region used for overlay analysis. (c) Overlay of the selected region showing co-registered EPMA and SEM-CL data, with machine-learning–assisted feature identification used to segment CL-dark features (red) and highlight candidate precipitate-associated regions (blue); arrows indicate examples where segmented CL-dark features co-localize with Fe-enriched precipitate signals. Red dashed circles show unmatched CL dark features that lack a corresponding Fe-rich precipitate. Green dashed circles show unmatched Fe-rich precipitates that lack a corresponding CL-dark feature. Scale bar corresponds to 5 $\mu$m.

To assemble a more objective comparison, we performed an overlay analysis on the full co-imaged field of view using IPSDK, a machine-learning-assisted feature-recognition program trained to segment CL-dark spots and identify candidate precipitate features. Then a multimodal registration approach using both affine and nonlinear transformation was performed on the overlay of CL-dark spots and EPMA Fe precipitate map. Although charging-induced drift in the insulating $ZrO_2$ introduced shear distortions that are poorly modeled by smooth polynomial warping during EBSD analysis, we found that the sparse distribution of Fe-precipitates allowed the CMA-ES algorithm to stabilize.

Across the full region, the algorithm identified 9 CL-dark spots overall, with only one on the right-hand side of the image. Therefore, Figure 5c focuses on the selected region of interest (orange boxes in Figures 5a and b) where numerous CL-dark features are recognized and overlap can be visually assessed. The overlay analysis within this region shows that 5 out of 8 of the CL-dark spots identified in SEM-CL co-localize with Fe-enriched features in EPMA (indicated with arrows in Figure 5c). Although EPMA also provides Cr maps, we did not pursue a rigorous Cr-CL co-localization analysis since Cr distributions are more uniformly associated with the original SPP footprint and less likely to be the cause of increased concentrations of CL dark spots near the metal-oxide interface. In contrast, Fe is an established CL quencher in many oxide/mineral systems [25] [26] [45] and exhibits stronger spatial heterogeneity in the oxide. Furthermore, the potential metallic nature of these Fe-rich precipitates may provide efficient carrier sinks, further enhancing the quenching effect. Overall, these results support the interpretation that a substantial fraction of the CL-dark features near the metal-oxide interface are associated with Fe precipitates, while also demonstrating that CL contrast is not purely chemical in origin. Together, the EBSD and EPMA correlations show that SEM-CL contrast in thermally grown $ZrO_2$ reflects spatial variations in carrier recombination, and thus transport, with systematic suppression near Fe enriched SPPs and in grain boundary rich or locally disordered regions.

## 4. Discussion

In this work we demonstrate cross-sectional SEM-CL microscopy of thermally grown zirconia on zirconium alloys and establish quantitative links between luminescence contrast and chemical and microstructural heterogeneities. The measured CL signal is dominated by the ~2.7 eV oxygen vacancy defect band [46] and exhibits spatial modulation across the oxide cross section. Co-registered EBSD and SEM-CL analysis shows that CL intensity decreases in regions with higher local orientation spread and grain-boundary-rich striations, consistent with enhanced non-radiative recombination associated with microstructural disorder. EPMA mapping of the same region shows Fe-rich features identified by EPMA systematically coincide with CL-dark spots, and these Fe-linked quenched regions concentrate in the first ~10 μm from the metal/oxide interface.

The chemical origin of CL suppression is consistent with non-radiative energy transfer by Fe centers [47] [48] and with the depth evolution expected for Fe-rich SPPs dissolution into the growing oxide [49] [40]. In short, SEM-CL serves as a practical proxy map of Fe-bearing precipitate fields and the associated electronic-trap landscape through the oxide thickness. Because Fe(II/III) centers and Fe-rich inclusions introduce mid-gap states and locally pin the Fermi level [8], they open efficient non-radiative recombination channels, which quench the luminescence otherwise produced by oxygen vacancy-related emitters present in the n-type $ZrO_2$ [3] [4]. Notably, this quenching mechanism does not depend on the specific Fe valence: in Fe-doped zirconia, suppression of the oxygen-vacancy emission persists when surface $Fe^{3+}$ is reduced to $Fe^{2+}$ without structural changes to PL [50]. Hence, the observed association of CL-dark features with Fe-rich precipitates is robust to the local valence distribution. The resulting CL-intensity minima therefore mark regions of high trap density and electronic band bending around SPPs, providing a mesoscale, field-of-view diagnostic of where electrons are most likely to be captured and where charge transport (and thus corrosion-relevant electronic flux) will be spatially biased. It is also clear that the CL signal tracks microstructural disorder. EBSD-CL correlation on matched regions shows an inverse relationship between mean orientation spread (a proxy for local texture heterogeneity) and total CL intensity: banded/columnar zones and grain-boundary-rich striations imaged by EBSD are distinctly dimmer in CL, consistent with enhanced non-radiative recombination at dislocations/GBs that depress the local quasi-Fermi level and shorten carrier lifetimes. Thus, beyond chemistry, the CL map records the spatial distribution of electronic recombination centers tied to the oxide crystallographic texture. These electronic-structure contrasts map directly onto the transport pathways that govern zirconium-alloy corrosion in light-water-reactor environments. In particular, oxide grain boundaries are preferential sites for nanopore formation, and nanopore density decreases as monoclinic oxide grains become more columnar and as a strong $\{10\bar{n}\}$ fiber texture develops [51] [52] [53]. Nanopore channels have also been shown to align with the oxide-growth direction, implying viable diffusion channels for oxidizing species [54] [55] [56]. In our study, we semi-quantitatively show that regions of heterogeneous local texture, which are known to contain a higher density of nanopores, are CL-quenching sites. Consequently, zones with high pore density/interconnectivity appear as CL-dark bands, linking the optical readout to the heterogeneous mass- and charge-transport pathways that govern pre-transition kinetics and the onset of oxide transition [57].

The transition from a columnar inner oxide to an equiaxed-appearing outer scale is consistent with established Zr-oxidation kinetics. This morphology reflects the evolution from a 'fossil' oxide, formed during the initial fast corrosion stage, to a more stable diffusion-limited growth regime. Additional EBSD-based texture and local misorientation analyses supporting the observed microstructural evolution across the oxide scale are provided in Section S2 of the Supplementary.

Finally, the SEM-CL signatures we observe (namely Fe-precipitate-induced and GB/texture-linked quenching) map onto microstructural factors long implicated in Zr-alloy corrosion. Although SPPs are known to influence corrosion rate, the causal pathway has remained relatively elusive. It is worth noting that measurable rate changes occur at SPP volume fractions of order ~1%, far below any percolation threshold, implying a long-range, field-mediated effect on charge transport rather than short-circuit transport through a connected SPP network, as often proposed in the literature [47] [58] [59] [60] [61]. Our correlative SEM-CL data, while semi-quantitative, support this view. Fe-rich precipitates, such as $Zr(Fe,Cr)_2$ Laves phases, impose a steady-state space-charge pattern in $ZrO_2$, locally pinning the Fermi level and introducing mid-gap traps that manifest as CL quenching. The resultant band bending and trap-assisted recombination bias electron/ion fluxes steering where charge is transported. While modeling work would need to be performed to measure the band-bending induced space charge layer thickness around the SPP, it could span tens to hundred nanometers, although this would depend on the interfacial density of states and dopant concentration in $ZrO_2$. In this framework, SPPs act as electrostatic/chemical sources, or local electrodes, that shape the oxide's electronic landscape well beyond their physical footprint, while grain-boundary/texture heterogeneity modulates that landscape via defect density and porosity. While SEM-CL resolution is not high enough to map the electronic structure at the nanoscale, it provides an electronic map of these long-range effects, identifying low-luminescence (high-trap) corridors predisposed to alter electronic/ionic transport.

## 5. Conclusion

In summary, this work demonstrates cross-sectional SEM-CL correlation microscopy as a sensitive mesoscale probe of electronic-defect heterogeneity in thermally grown zirconia on zirconium alloys. By co-registering SEM-CL with EBSD and EPMA on the same oxide cross-section, we establish that the CL response, dominated by the ~2.7 eV defect band, is not spatially uniform, but instead exhibits systematic modulation that can be quantitatively linked to microstructural and chemical features. In particular, the correlative EBSD-CL analysis shows that CL intensity increases with grain area and decreases with grain-boundary-rich and locally disordered regions, consistent with enhanced non-radiative recombination associated with microstructural disorder. These observations highlight that SEM-CL contrast cannot be interpreted solely as a thickness or morphology effect in corrosion-grown zirconia, rather, it directly reflects spatial variations in recombination pathways that are governed by the evolving oxide microstructure. In addition, co-registered EPMA mapping reveals that a substantial fraction of CL-dark features co-localize with Fe-enriched secondary phase precipitates embedded in the inner oxide, supporting a chemically driven quenching mechanism in which metallic inclusions and/or dissolved Fe centers act as efficient carrier sinks and non-radiative recombination sites.

The combined datasets therefore demonstrate that SEM-CL encodes both microstructure-driven and chemistry-driven suppression of radiative recombination, underscoring the necessity of multimodal registration to disentangle competing contrast mechanisms in zirconia. More broadly, this correlative framework provides an efficient route to connect luminescence signatures to underlying defect landscapes in heterogeneous oxides, offering a pathway to map regions most likely to bias electronic/ionic transport during oxidation and, by extension, to inform mechanistic understanding of oxide growth and degradation in high-temperature energy. Extending this correlative approach to a series of samples with systematically varied alloy compositions and oxidation conditions is a natural next step toward quantitatively linking luminescence signatures to corrosion behavior.

## Acknowledgement

This work was primarily supported by the U.S. Department of Energy, Office of Science, Basic Energy Sciences under Award DE-SC0023873, with the initial SEM-CL characterization supported through a seed grant by the NSF Division of Materials Research through the University of Wisconsin Materials Research Science and Engineering Center (Grant No. DMR-1720415). The authors gratefully acknowledge use of facilities and instrumentation in the UW-Madison Wisconsin Center for Nanoscale Technology. The Center (wcnt.wisc.edu) is partially supported by the Wisconsin Materials Research Science and Engineering Center (NSF DMR-2309000) and the University of Wisconsin-Madison.

## References

[1] H. Kania, "Corrosion and anticorrosion of alloys/metals: the important global issue," *Coatings,* vol. 13, no. 2, p. 216, 2023.

[2] J. Liu, A. H. Mir, G. He, M. Danaie, J. Hinks, S. Donnelly, H. Nordin, S. Lozano-Perez and C. R. Governor, "In-situ TEM study of irradiation-induced damage mechanisms in monoclinic-ZrO2," *Acta Materialia,* vol. 199, pp. 429-442, 2020.

[3] H. K. Yueh and B. Cox, "Cathodoluminescence imaging of oxdised zirconium alloys," *Journal of Nuclear Materials,* vol. 324, no. 2-3, pp. 203-214, 2004.

[4] H. K. Yueh and B. Cox, "Luminescence properties of zirconium oxide films," *Journal of Nuclear Materials,* vol. 323, no. 1, pp. 57-67, 2003.

[5] L. Grigorjeva, D. Millers, A. Kalinko, V. Pankratov and K. Smits, "Time-resolved cathodoluminescence and photoluminescence of nanoscale oxides," *Journal of the European Ceramic Society,* vol. 29, no. 2, pp. 255-259, 2009.

[6] A. Martinez-Hernandez, J. Guzman-Mendoza, T. Rivera-Montalvo, D. Sanchez-Guzman, J. Guzman-Olguin, M. Garcia-Hipolito and C. Falcony, "Synthesis and cathodoluminescence characterization of ZrO2:Er3+ films," *Jounral of Luminescence,* vol. 153, pp. 140-143, 2014.

[7] M. Boffelli, W. Zhu, M. Back, G. Sponchia, T. Francese, P. Riello, A. Benedetti and G. Pezzotti, "Oxygen hole states in Zirconia Lattices: Quantitative Aspects of their Cathodoluminescence Emission," *The Journal of Physical Chemistry,* vol. 118, no. 42, pp. 9828-9836, 2014.

[8] R. Benaboud, P. Bouvier, J.-P. Petit, Y. Wouters and A. Galerie, "Comparative study and imaging by PhotoElectroChemical techniques of oxide films thermally grown on zirconium and Zircaloy-4," *Journal of Nuclear Materials,* vol. 360, no. 2, pp. 151-158, 2007.

[9] K. Smits , L. Grigorjeva, D. Millers, A. Sarakovskis, J. Grabis and W. Lojwoski, "Intrinsic defect related luminescence in ZrO2," *Journal of luminescence,* vol. 131, no. 10, pp. 2058-2062, 2011.

[10] T. Perevalov, D. Gulyaev, V. Aliev, K. Zhuravlev, V. Gritsenko and A. Yelisseyev, "The origin of 2.7 eV blue luminescence band in zirconium oxide," *Journal of Applied Physics,* vol. 116, no. 24, p. 244109, 2014.

[11] J. Ramos-Justicia, J. Ballester-Andújar, A. Urbieta and P. Fernández, "Growth of Zr/ZrO2 Core--Shell Structures by Fast Thermal Oxidation," *Applied Sciences,* vol. 13, p. 3714, 2023.

[12] R. Hahn, S. Berger and P. Schmuki, "Bright visible luminescence of self-organized ZrO2 nanotubes," *Journal of solid state electrochemistry,* vol. 14, pp. 285-288, 2010.

[13] J. Hu, A. Garner, N. Ni, A. Gholinia, R. J. Nicholls, S. Lozano-Perez, P. Frankel, M. Preuss and C. R. Grovenor, "Identifying suboxide grains at the metal-oxide interface of a corroded Zr-1.0%Nb alloy using (S)TEM, transmission-EBSD, and EELS," *Micron,* vol. 69, pp. 35-42, 2015.

[14] A. Baris, R. Restani, R. Grabherr, Y.-L. Chiu, H. Evans, K. Ammon, M. Limback and S. Abolhassani, "Chemical and microstructural characterization of a 9 cycle Zircaloy-2 cladding using EPMA and FIB tomography," *Journal of Nuclear Materials,* vol. 504, pp. 144-160, 2018.

[15] P. R. Edwards and R. W. Martin, "Cathodoluminescence nano-characterization of semiconductors," *Semiconductor Science and Technology,* vol. 26, no. 6, p. 064005, 2011.

[16] A. Gustafsson and E. Kapon, "Cathodoluminescence in the scanning electron microscope: application to low-dimensional semiconductor structures," *Scanning Mirosc,* vol. 12, no. 2, pp. 285-299, 1998.

[17] T. Duong, H. K. Mulmudi, H. Shen, Y. Wu, C. Barugkin, Y. O. Mayon, H. T. Nguyen, D. Macdonald, J. Peng, M. Lockrey, W. Li, Y.-B. Cheng, T. P. White, K. Weber and K. Catchpole, "Structural engineering using rubidium iodide as a dopant under excess lead iodide conditions for high efficiency and stable perovskites," *Nano Energy,* vol. 30, pp. 330-340, 2016.

[18] D. Cortecchia, K. C. Lew, J.-K. So, A. Bruno and C. Soci, "Cathodoluminescence of Self-Organized Heterogenous Phases in Multidimensional Perovskite Thin Films," *Chemistry of Materials,* vol. 29, no. 23, pp. 10088-10094, 2017.

[19] J. Moseley, M. M. Al-Jassim, D. Kuciauaskas, H. R. Moutinho, N. Paudel, H. L. Guthrey, Y. Yan, W. K. Metzger and R. K. Ahrenkiel, "Cathodoluminescence Analysis of Grain Boundaries and Grain Interiors in Thin-Film CdTe," *IEEE Journal of Photovoltaics,* vol. 4, no. 6, pp. 1671-1679, 2014.

[20] J. Moseley, P. Rale, S. Collin, E. Colegrove, H. Guthrey, D. Kuciauskas, H. Moutinho, M. Al-Jassim and W. K. Metzger, "Luminescence methodology to determine grain-boundary, grain-interior, and surface recombination in thin-film solar cells," *Journal of Applied Phyiscs,* vol. 124, no. 11, 2018.

[21] S. N. Frelinger, M. D. Ledvina, J. R. Kyle and D. Zhao, "Scanning electron microscopy cathodoluminescence of quartz: Principles, techniques and applications in ore geology," *Ore Geology Reviews,* vol. 65, pp. 840-852, 2015.

[22] W. P. Leeman, C. M. MacRae, N. C. Wilson, A. Torpy, C.-T. A. Lee, J. J. Student, J. B. Thomas and E. P. Vicenzi, "A study of cathodoluminescence and trace element compositional zoning in natural quartz from volcanic rocks: mapping titanium content in quartz," *Microscopy and microanalysis,* vol. 18, no. 6, pp. 1322 - 1341, 2012.

[23] T. Sekiguchi, J. Hu and Y. Bando, "Cathodoluminescence study of one-dimensional free-standing widegap-semiconductor nanostructures: GaN nanotubes, Si3N4 nanobelts and ZnS/Si nanowires," *Journal of electron microscopy,* vol. 53, no. 2, pp. 203-208, 2004.

[24] L. Blaine, Hydrothermal fluids and Cu-Au mineralization of the deep grasberg porphyry deposit, papua, indonesia, Austin: The University of Texas at Austin, 2007.

[25] N. Galili, I. Kaplan-Ashiri and I. Halevy, "Cathodoluminescence of iron oxides and oxyhydroxides," *American MIneralogist,* vol. 108, no. 8, pp. 1436-1448, 2023.

[26] D. A. Budd, U. Hammes and W. B. Ward, "Cathodoluminescence in calcite cements: new insights on Pb and Zn sensitizing, Mn activation, and Fe quenching at low trace-element concentrations," *Journal of sedimentary research,* vol. 70, no. 1, pp. 217-226, 2000.

[27] C. Ballesteros and J. LLopis, "Cathodoluminescence from deformed doped MgO crsytals," *Radiation Effects,* vol. 74, no. 1-4, pp. 347-351, 1983.

[28] M. Hamers, G. Pennock and M. Drury, "Scanning electron microscope cathodoluminescence imaging of subgrain boundaries, twins, and planar deformation features in quartz," *Phys Chem Minerals,* vol. 44, pp. 263-275, 2017.

[29] A. Couet, L. Borrel, J. Liu, J. Hu and C. Grovernor, "An integrated modeling and experimental approach to study hydrogen pickup mehcanism in zirconium alloys," *Corrosion Science,* vol. 159, p. 108134, 2019.

[30] J. Romero, J. Partezana, R. Comstock, L. Hallstadius, A. Motta and A. Couet, "Evolution of hydrogen pickup fraction with oxidation rate on zirconium alloys," presented at TopFuel 2015 - unpublished, Zurich, Sep. 13–17, 2015.

[31] A. T. Motta, A. Couet and R. J. Comstock, "Corrosion of zirconium alloys used for nuclear fuel cladding," *Annual Review of Materials Research,* vol. 45, no. 1, pp. 311-343, 2015.

[32] A. Yilmazbayhan, M. Gomes, A. Motta, H.-G. Kim, Y. H. Jeong, J.-Y. Park, R. Comstock, B. Lai and Z. Cai, "Characterization of oxides formed on model zirconium alloys in 360 c water using micro-beam synchrotron radiation," *Proceedings of the 12th International Conference on Environmental Degradation of Materials in Nuclear Power System--Water Reactors, Salt Lake City, UT,* p. 201, 2005.

[33] J. Götze, M. Plötze and D. Habermann, "Origin, spectral characteristics and practical applications of the cathodoluminescence (CL) of quartz - a review," *Mineralogy and petrology,* vol. 71, pp. 225-250, 2001.

[34] D. Dingley, "Progressive steps in the development of electron backscatter diffraction and orientation imaging microscopy," *Journal of microscopy,* vol. 213, pp. 214-224, 2004.

[35] J. J. Donovan and T. N. Tingle, "An improved mean atomic number background correction for quantitative microanalysis," *Microscopy and Microanalysis,* vol. 2, pp. 1-7, 1996.

[36] D. Drouin, A. Couture, D. Joly, X. Tastet, V. Aimez and R. Gauvin, "CASINO V2. 42—a fast and easy-to-use modeling tool for scanning electron microscopy and microanalysis users," *Scanning: The Journal of Scanning Microscopies,* vol. 29, pp. 92-101, 2007.

[37] T. Wang, G. Wang, M. Qiu, W. Cheng, J. Zhang and G. Zhao, "The origin of the 500 nm luminescence band related to oxygen vacancies in ZrO2," *Journal of Luminescence,* vol. 237, p. 118133, 2021.

[38] H. Beck and C. Kaliba, "On the solubility of Fe, Cr, and Nb in ZrO2 and its effect on thermal dilatation and polymorphic transition," *Materials Research Bulletin,* vol. 25, no. 9, pp. 1161-1168, 1990.

[39] M.-A. Charpagne, F. Strub and T. M. Pollock, "Accurate reconstruction of EBSD datasets by a multimodal data approach using an evolutionary algorithm," *Materials Characterization,* vol. 150, pp. 184-198, 2019.

[40] D. Pecheur, F. Lefebvre, A. Motta, C. Lemaignan and J. Wadier, "Precipitate evolution in the Zircaloy-4 oxide layer," *Journal of Nuclear Materials,* vol. 189, no. 3, pp. 318-332, 1992.

[41] K. Annand, M. Nord, I. MacLaren and M. Gass, "The corrosion of Zr (Fe, Cr) 2 and Zr2Fe secondary phase particles in Zircaloy-4 under 350 C pressurised water conditions," *Corrosion Science,* vol. 128, pp. 213-223, 2017.

[42] J. Hu, B. Setiadinata, T. Aarholt, A. Garner, A. Vilalta-Clemente, J. Partezana, P. Frankel, P. Bagot, S. Lozano-Perez and A. e. a. Wilkinson, "Understanding Corrosion and Hydrogen Pickup of Zirconium Fuel Cladding Alloys: The Role of Oxide Microstructure," *Porosity, Suboxides, and Second-Phase Particles, Zirconium in the Nuclear Industry,* pp. 93-126, 2018.

[43] C. R. Grovenor, N. Ni, D. Hudson, S. S. Yardley, K. L. Moore, G. D. Smith, S. Lozano-Perez and J. M. Sykes, "Mechanisms of oxidation of fuel cladding alloys revealed by high resolution APT, TEM and SIMS analysis," *MRS Online Proceedings Library (OPL),* vol. 1383, pp. mrsf11-1383, 2012.

[44] A. Couet, A. T. Motta, B. De Gabory and Z. Cai, "Microbeam X-ray absorption near-edge spectroscopy study of the oxidation of Fe and Nb in zirconium alloy oxide layers," *Journal of Nuclear Materials,* vol. 452, pp. 614-627, 2014.

[45] J. Sima, "(Non)luminescent properties of iron compounds," *Acta Chimica Slovaca,* vol. 8, no. 2, pp. 126-132, 2015.

[46] J. Robertson, K. Xiong and S. Clark, "Band gaps and defect levels in functional oxides," *Thin Solid Films,* vol. 496, pp. 1-7, 2006.

[47] Y. Hatano, K. Isobe, R. Hitaka and M. Sugisaki, "Role of intermetallic precipitates in hydrogen uptake of Zircaloy-2," *Journal of nuclear science and technology,* vol. 33, pp. 944-949, 1996.

[48] X. Iltis, F. Lefebvre and C. Lemaignan, "Microstructural study of oxide layers formed on Zircaloy-4 in autoclave and in reactor part 11: Impact of the chemical evolution of intermetallic precipitates on their zirconia environment," *Journal of nuclear materials,* vol. 224, pp. 121-130, 1995.

[49] J. H. Baek and Y. H. Jeong, "Depletion of Fe and Cr within precipitates during Zircaloy-4 oxidation," *Journal of nuclear materials,* vol. 304, pp. 107-116, 2002.

[50] H. Nakajima, K. Itoh, H. Kaneko and Y. Tamaura, "Effects of Fe doping on crystalline and optical properties of yttria-stabilized zirconia," *Journal of Physics and Chemistry of Solids,* vol. 68, pp. 1946-1950, 2007.

[51] H. Zhang, R. Su, B. Queylat, T. Kim, G. Lucadamo, W. Howland and A. Couet, "3D reconstruction and interconnectivity quantification of the nano-porosity in the oxide layer of corroded Zr alloys," *Corrosion Science,* vol. 226, p. 111630, 2024.

[52] H. Zhang, T. Kim, J. Swarts, Z. Yu, R. Su, L. Liu, W. Howland, G. Lucadamo and A. Couet, "Nano-porosity effects on corrosion rate of Zr alloys using nanoscale microscopy coupled to machine learning," *Corrosion Science,* vol. 208, p. 110660, 2022.

[53] M. S. Yankova, A. Garner, F. Baxter, S. Armson, C. P. Race, M. Preuss and P. Frankel, "Untangling competition between epitaxial strain and growth stress through examination of variations in local oxidation," *Nature Communications,* vol. 14, p. 250, 2023.

[54] A. Garner, A. Gholinia, P. Frankel, M. Gass, I. MacLaren and M. Preuss, "The microstructure and microtexture of zirconium oxide films studied by transmission electron backscatter diffraction and automated crystal orientation mapping with transmission electron microscopy," *Acta Materialia,* vol. 80, pp. 159-171, 2014.

[55] J. Hu, T. Aarholt, B. Setiadinata, K. Li, A. Garner, S. Lozano-Perez, M. Moody, P. Frankel, M. Preuss and C. Grovenor, "A multi-technique study of “barrier layer” nano-porosity in Zr oxides during corrosion and hydrogen pickup using (S) TEM, TKD, APT and NanoSIMS," *Corrosion Science,* vol. 158, p. 108109, 2019.

[56] N. Ni, S. Lozano-Perez, M. Jenkins, C. English, G. Smith, J. Sykes and C. Grovenor, "Porosity in oxides on zirconium fuel cladding alloys, and its importance in controlling oxidation rates," *Scripta Materialia,* vol. 62, pp. 564-567, 2010.

[57] B. Ensor, A. Motta, A. Lucente, J. Seidensticker, J. Partezana and Z. Cai, "Investigation of breakaway corrosion observed during oxide growth in pure and low alloying element content Zr exposed in water at 360° C," *Journal of Nuclear Materials,* vol. 558, p. 153358, 2022.

[58] Y. Hatano, R. Hitaka, M. Sugisaki and M. Hayashi, "Influence of size distribution of Zr (Fe, Cr) 2 precipitates on hydrogen transport through oxide film of Zircaloy-4," *Journal of nuclear materials,* vol. 248, pp. 311-314, 1997.

[59] B. Cox, "A mechanism for the hydrogen uptake process in zirconium alloys," *Journal of nuclear materials,* vol. 264, pp. 283-294, 1999.

[60] B. Cox and Y.-M. Wong, "A hydrogen uptake micro-mechanism for Zr alloys," *Journal of nuclear materials,* vol. 270, pp. 134-146, 1999.

[61] B. Cox, "Some thoughts on the mechanisms of in-reactor corrosion of zirconium alloys," *Journal of Nuclear materials,* vol. 336, pp. 331-368, 2005.

**Supplementary Materials for Mapping optical, chemical, structural features in $ZrO_2$ via cross-sectional SEM-Cathodoluminescence correlation microscopy**

Ricardo Vidrio[1*], Yuhan Tong[1*], Junliang Liu[4], Bil Schneider[5], William O. Nachlas[5], Nathan Curtis[2], Maryam Zahedian[1], Alexander Kvit[3], Hongliang Zhang[3], Zefeng Yu[2], Ximeng Wang[2], Yongfeng Zhang[2], Adrien Couet[2], and Jennifer T. Choy[1]

[1]Department of Electrical and Computer Engineering, University of Wisconsin-Madison, United States

[2]Department of Nuclear Engineering and Engineering Physics, University of Wisconsin-Madison, United States

[3]Department of Material Science and Engineering, University of Wisconsin-Madison, United States

[4]Department of Materials Science and Engineering, FAMU-FSU College of Engineering, Florida State University, United States

[5]Department of Geoscience, University of Wisconsin-Madison, United States

* These authors contributed equally to this work.

This document contains:

- Inverse Pole Figure (IPF) maps collected in EBSD for $ZrO_2$ monoclinic and zirconia tetragonal grains
- Comparison of Pole Figure (PF) maps for ZrO2 monoclinic in its outer and inner region
- Comparison of Misorientation Distribution Function (MDF) maps for ZrO2 monoclinic in its outer and inner region
- ROI selection for EBSD comparison with SEM-CL

1. IPF analysis

We performed an IPF analysis to quantify grain orientations in the EBSD data and investigate how crystallographic texture influences cathodoluminescence contrast and the phase transformation behavior in the oxide layer.

The Inverse Pole Figure (IPF) $Y_1$ (nominal sample normal to M/O interface) map for monoclinic $ZrO_2$ (Figure S1 a) shows clustering of grains mostly around {001} and {-100} directions, indicative of a fiber texture. Grains contributing to this fiber texture have their crystallographic axes preferentially aligned along $Y_1$ and with minimal contribution from the {010} axis. The relatively random color distribution in the $X_1$ (transverse direction) and $Z_1$ (out of page direction) maps further supports a fiber-like texture along $Y_1$.

In the tetragonal zirconia map (Figure S1 b), the grains display a specific clustering with the {001} axis preferentially aligned along the $X_1$ direction of the sample. Another cluster, corresponding to the {010} axis, aligns near the $Y_1$ and $Z_1$ direction, which suggests partial orientation relationships within the tetragonal zirconia.

The minor tetragonal component shows a sharper texture compared to the monoclinic phase. This observation is consistent with the tetragonal phase acting as the parent structure, from which monoclinic variants nucleate, likely through a transformation-toughening mechanism [34]. Such preferential variant selection suggests that the texture of the parent phase strongly influences the resulting monoclinic grain orientations.

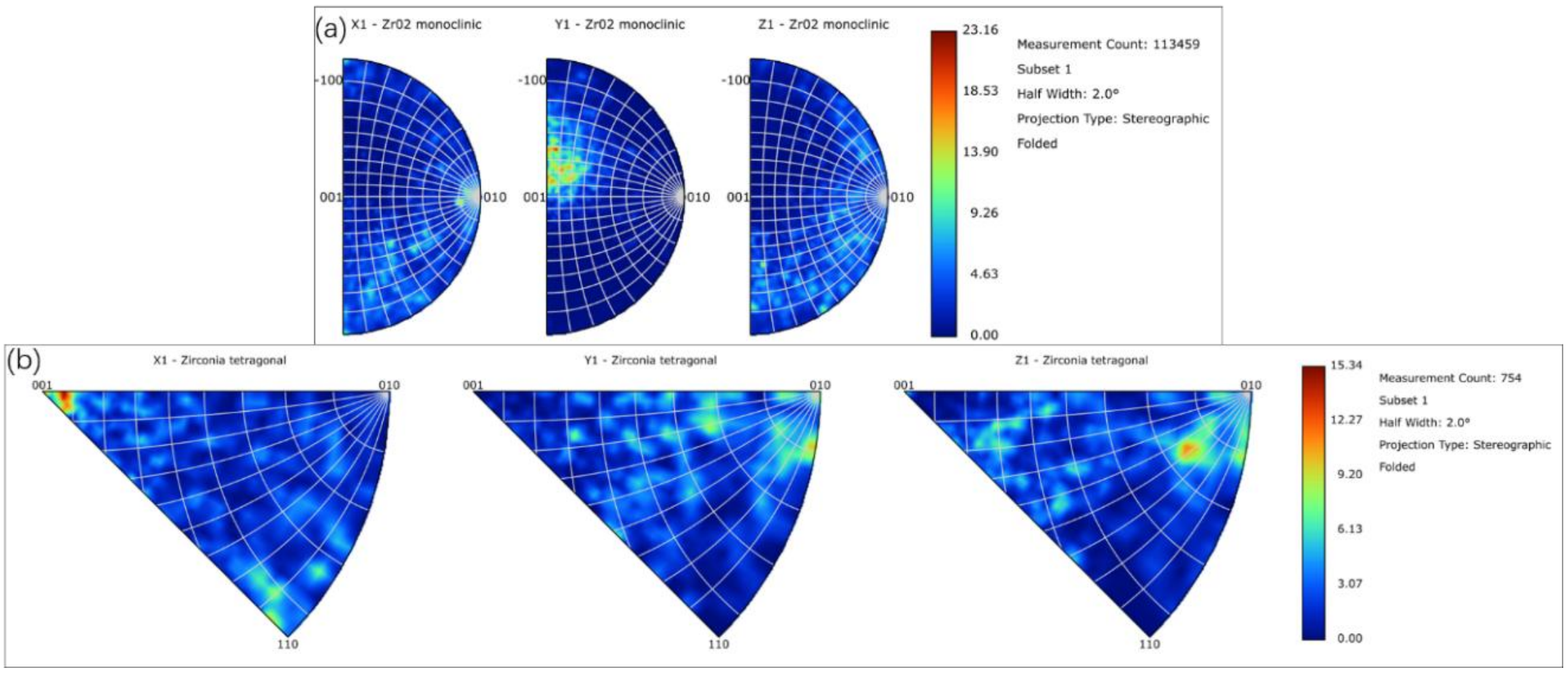

Figure S1: IPF maps plotted on Aztec Crystal software. a) $ZrO_2$ monoclinic and b) zirconia tetragonal IPF on $X_1$ (EBSD's image's horizontal to the right direction), $Y_1$ (vertical to the up direction), and $Z_1$ (out of page).

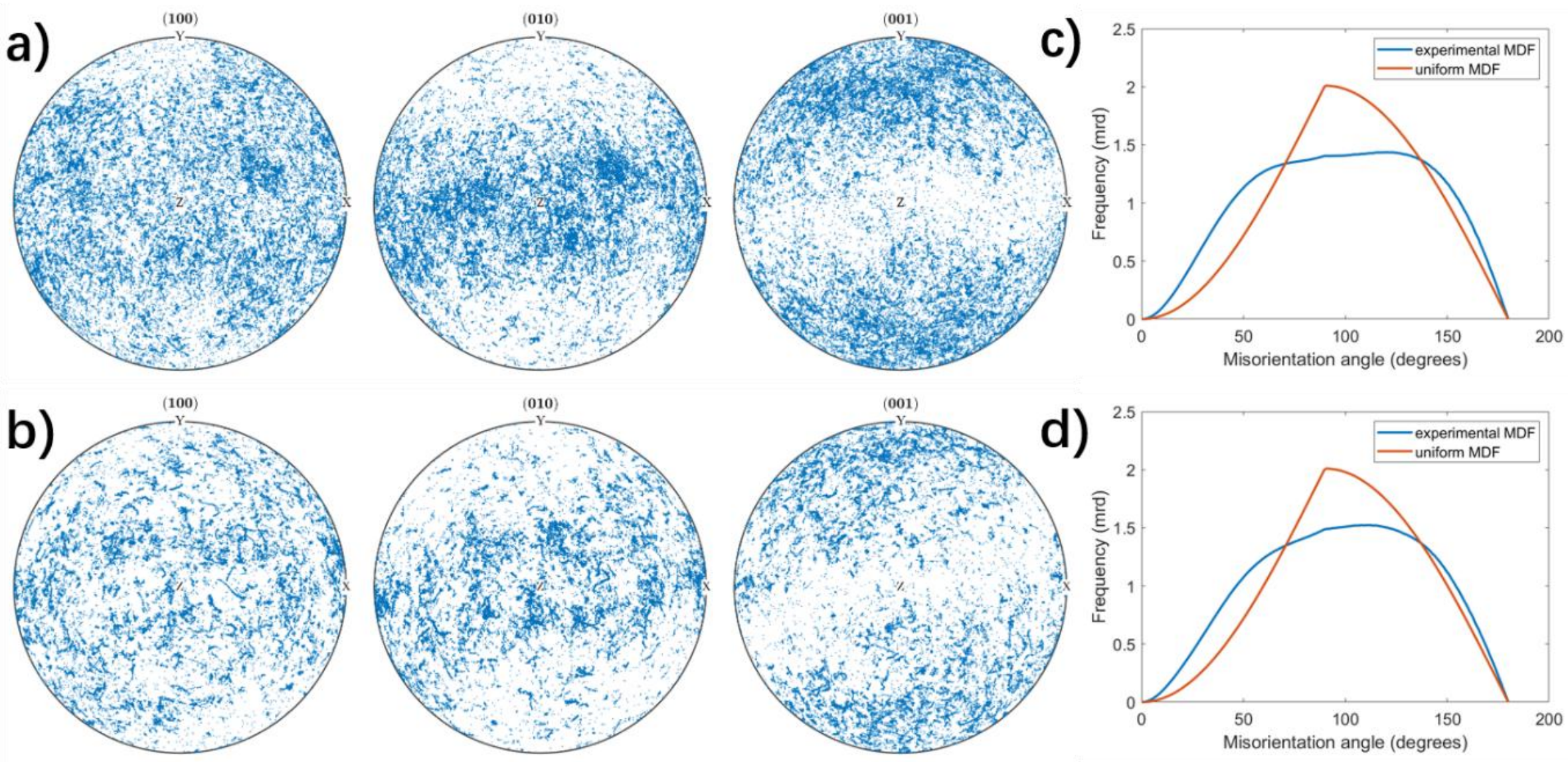

Figure S2: a) Outer region and b) inner region of ZrO2 monoclinic pole figure maps for {100}, {010}, and {001} planes plotted using MTEX toolbox [35] on MATLAB. c) Outer region and d) inner region of ZrO2 monoclinic Misorientation distribution maps (MDFs) [36] comparing the experimental data (blue) and a theoretical random texture (uniform MDF, red).

2. Regime-specific PF and MDF analysis

While the Inverse Pole Figure (IPF) maps provide a spatial overview of the fiber-like texture across the oxide scale, a more quantitative analysis was conducted to compare the inner region (~0–10 μm from the interface) and the outer region (~10–30 μm). This separation allows us to evaluate the evolution in texture in the context of the Fe and Cr precipitate distribution identified by EPMA.

Pole figures (Figure 2a, b) were generated for both regions to identify the shifts in orientation distribution. Both layers showed a preference for the {010} poles to align near the center (X and Z sample axes), and for {001} poles to align near the Y axis (growth direction). However, the inner layer exhibits high orientation scatter with discrete crystallographic clusters. This spatially coincides with the presence of metallic precipitates near the interface, contrasting with the smoother orientation distribution observed in the outer layer where precipitates have largely oxidized.

To quantify the grain boundary statistics, the MDF was calculated for both inner and outer monoclinic layers (Figure 2c, d). While both regions confirm a preferred crystallographic orientation rather than a fully randomized state, the inner layer distribution is closer to the uniform (random) curve.

This finding is consistent with a transition in oxidation kinetics. In the inner region, where growth dominates over nucleation, the grains are characteristically larger and columnar. However, the proximity of metallic $Zr(Fe,Cr)_2$ particles at this growth front is associated with a more randomized orientation distribution.

As these SPPs move toward the inner region, they undergo a well-studied oxidation pathway characterized by Fe out diffusion and the formation of Cr-O remnants of the outer region. As the precipitates oxidize, the grains appear to achieve the stronger divergence from the uniform MDF observed in the surface-proximate layers.

By comparing the Mean Orientation Spread (MOS, in Figure 4) and MDF, we can see opposite trends across the scale thickness. The outer region exhibits a higher MOS alongside its more divergent MDF. While the MDF reflects a more organized collective texture (inter-granular alignment), the high MOS indicates significant intra-granular lattice curvature and accumulated growth-induced strain.

## 3. ROI analysis

To evaluate the influence of crystallographic structures on cathodoluminescence in SEM-CL data, spatial correlation was performed and the images of EBSD maps and SEM-CL intensity maps were aligned.

A 9-region (3x3) grid sampling method (Figure 3a) was employed to ensure statistical representation across the ROI. Then each grid's EBSD image is binarized, as well as the SEM-CL image, for the multimodal registration method's application (Github code forked and adapted from Charpagne et al. [30] https://github.com/yuhantong177/distortions). Alignment was achieved using a modified affine transformation (scaling, translation, and rotation) [30]. Prominent landmark grains visible in both modalities (in grid 1) were used as anchors (Figure S3 c). To prevent unphysical inconsistencies in rotation/translation between grids, the affine transformation adjusted using the anchor was applied to all sub-grids simultaneously rather than individual local alignments (Figure S3 d). Then every corresponding grid on the 8-bit grayscale SEM-CL image has CL intensity measured on ImageJ software from 0-255.

We then compared the EBSD grain information of each of the 9 subsets, examining the grain size, mean and maximum orientation spread, and number of grains per area, which are correlated against the mean CL intensity from SEM-CL's corresponding regions.

Figure S4 (a) shows a positive correlation between average grain area vs mean CL intensity. This is consistent with larger grains having fewer boundaries (which act as non-radiative recombination centers, which instead of emitting photons, releasing energy as phonons, decreasing CL intensity). (b) shows a negative correlation between mean orientation spread vs mean CL intensity. This means higher local misorientation within a grain (due to dislocations or strain etc.) corresponds to a dimmer area. Similar to 2, (c) shows that regions with higher luminescence area are crystallographically less defective. The negative correlation in (d) is expected since smaller grains means more boundaries which quench luminescence.

Notably, for the outlier in grid 1, it shows a higher mean CL intensity that deviates from the crystallographic trend. This is because the CL signal of this grid is dominated by the 2-3 large and bright grains in the grid. However, the EBSD statistics are averaged over ~100 total grains in the same grid, the influence of these giant grains is statistically diluted. Thus, the grid shows "average" grain statistics but skyrocketing CL intensity. Furthermore, we must account for the difference in signal excitation depths between the two techniques. The observed CL intensity may be influenced by large, high-intensity grains beneath the surface that are not captured in the EBSD map.

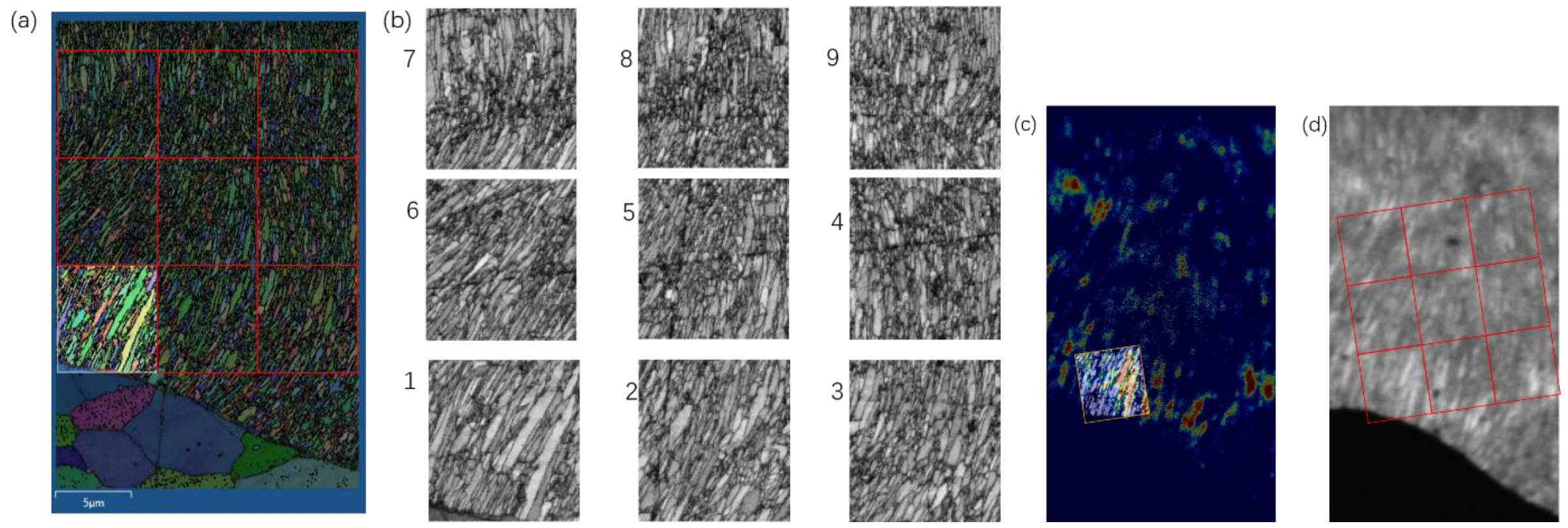

Figure S3: (a) EBSD map with band contrast overlay with grain boundaries, shown on Aztec Crystal. (b) EBSD subregions used for statistics corresponding to 9 data points in Figure 3. (c) EBSD grid 1 overlay onto SEM-CL speckle image. (d) all 3x3 grid of the SEM-CL data corresponding to the EBSD 9 subsets.

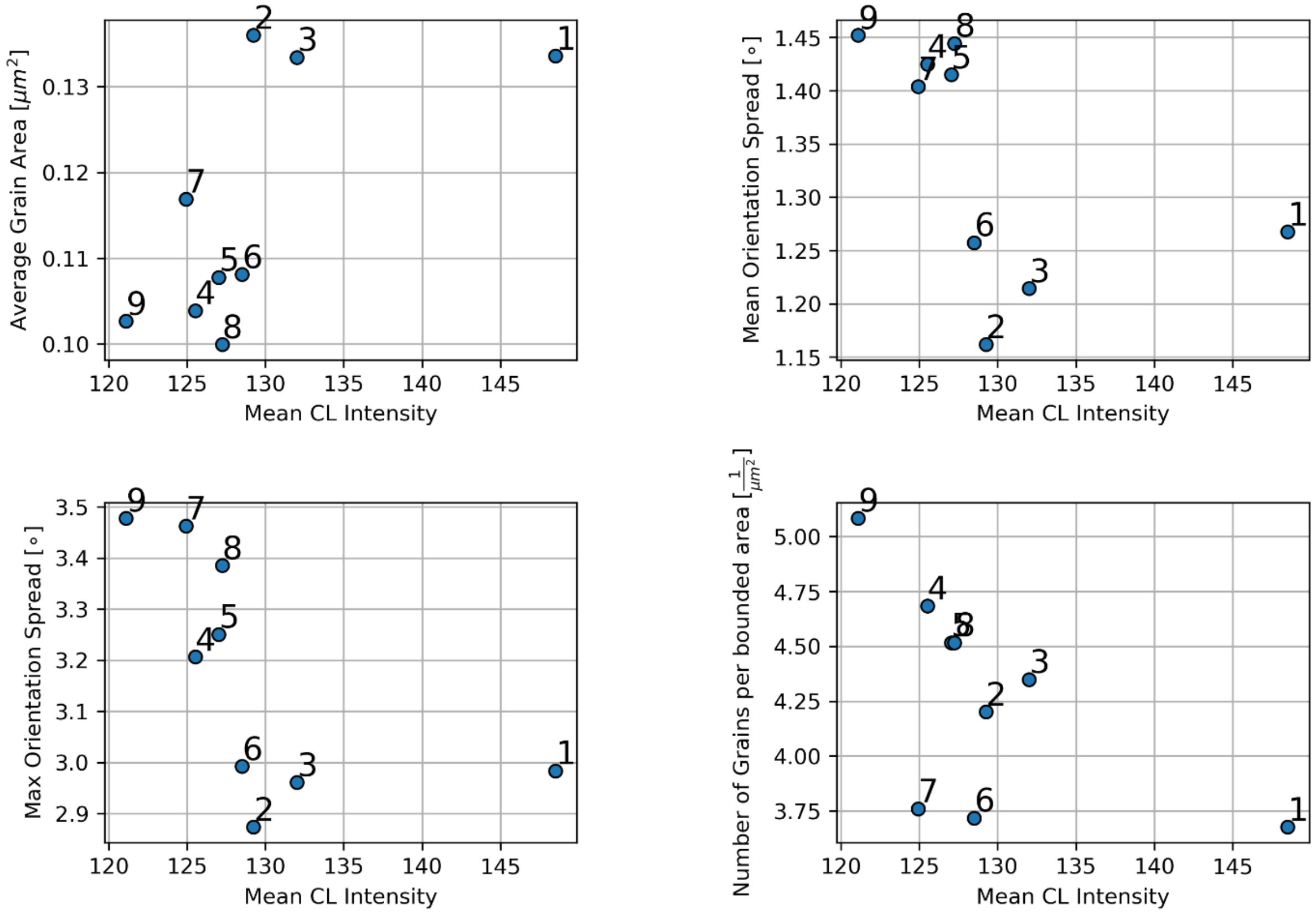

Figure 4: EBSD grain information from 9 regions correlated to the mean CL intensity from 9 corresponding regions of SEM-CL. The labels indicate which data point corresponds to which grid defined in Figure 3.

Reference:

[30] M.-A. Charpagne, F. Strub and T. M. Pollock, "Accurate reconstruction of EBSD datasets by a multimodal data approach using an evolutionary algorithm," *Materials Characterization,* vol. 150, pp. 184-198, 2019.

[34] Chevalier, Jerome, et al. "The tetragonal-monoclinic transformation in zirconia: lessons learned and future trends." *Journal of the american ceramic society* 92.9 (2009): 1901-1920.

[35] Hielscher, R., Schaeben, H., & Siemes, H. (2010). Orientation distribution within a single hematite crystal. *Mathematical Geosciences*, *42*(4), 359-375.

[36] Hielscher, Ralf, and Helmut Schaeben. "A novel pole figure inversion method: specification of the MTEX algorithm." *Applied Crystallography* 41.6 (2008): 1024-1037.